\documentclass[%
 reprint,
 amsmath,amssymb,
prb,
]{revtex4-2}

\usepackage[utf8]{inputenc}
\usepackage[T1]{fontenc}

\usepackage{graphicx}
\usepackage{dcolumn}
\usepackage{bm}
\usepackage{xcolor}
\usepackage[colorlinks=true,linkcolor=blue,citecolor=blue,urlcolor=blue]{hyperref}
\usepackage{physics}
\usepackage{natbib}
\begin{document}

\title{Spread of Correlations in Strongly Disordered Lattice Systems\\with Long-Range Coupling}

\author{Karol Kawa}
\email{karol.kawa@pwr.edu.pl}
\author{Pawe{\l} Machnikowski}%
\email{pawel.machnikowski@pwr.edu.pl}
\affiliation{Department of Theoretical Physics, Wroc{\l}aw University of Science and Technology, 50-370 Wroc{\l}aw, Poland}

\date{\today}

\begin{abstract}
We investigate the spread of correlations carried by an excitation in a 1-dimensional lattice system with high on-site energy disorder and long-range couplings with a power-law dependence on the distance ($\propto r^{-\mu}$).
The increase in correlation  between the initially quenched node and a given node exhibits three phases: quadratic in time, linear in time, and saturation. No further evolution is observed in the long time regime.
We find an approximate solution of the model valid in the limit of strong disorder and reproduce the results of numerical simulations with analytical formulas. We also find the time needed to reach a given correlation value as a measure of the propagation speed. Because of the triple phase evolution of the correlation function the propagation changes its time dependence. In the particular case of $\mu=1$, the propagation starts as a ballistic motion, then, at a certain crossover time, turns into standard diffusion.
\end{abstract}

\maketitle

\section{\label{sec:introduction}Introduction}

In relativistic physics, information cannot propagate faster than with the speed of light $c$. 
The upper limit for the propagation of information is determined by the so-called light cone, which means that the minimum time for sending information between points distant by $r$ is $t=r/c$.

Non-relativistic quantum theory does not impose any limitation on the speed of information propagation in the quantum system explicitly.
However, an upper limit of the speed at which quantum information can be transmitted can be induced by finite range interactions.
Indeed, Lieb and Robinson \cite{Lieb1972TheSystems, Robinson1976PropertiesSystems} proved the existence of such a limit using a lattice model with finite range interactions (decreasing at least exponentially).
They showed that for times $t < r/v$ the correlation between nodes distant by $r$ decreases exponentially.

The presence of this Lieb-Robinson boundary has been observed in many theoretical and experimental studies  \cite{Braganca2021QuenchLeads, Chougale2020DynamicsWaveguide,Cheneau2012, Langen2013LocalSystem, Hild2014Far-from-equilibriumMagnets, Richerme2014Non-localInteractions, Jurcevic2014QuasiparticleSystem, Lienhard2018, Despres2019TwofoldGaz,Carrega2021UnveilingFunctions}, e.g. the first experimental evidence was achieved in the system of a one-dimensional quantum gas trapped in an optical lattice \cite{Cheneau2012}.
Those authors focused on the evolution of the two-point correlation function after the local quench, tracking the time of the maximum correlation for successive atoms.
In this way, the maximum velocity of the correlation propagation was demonstrated.

While the Lieb-Robinson bound applies to locally interacting systems, much recent effort has been devoted to proving the existence of a similar limit in lattice systems with long-range interactions, decreasing  with the distance according to the power law, $\propto 1/r^\mu$ \cite{Kuwahara2020,Chen2019FiniteInteractions,Tran2020HierarchyInteractions, Hauke2013, Schachenmayer2014, Jurcevic2014QuasiparticleSystem, Richerme2014Non-localInteractions}.
The first mathematical evidence for the existence of a linear light cone in systems with long-range interactions was given in Ref.~\cite{Chen2019FiniteInteractions}. For $\mu>3$, they show that the time required to correlate distant atoms increases at least linearly with distance.
In Ref.~\cite{Kuwahara2020}, authors yield the mathematical evidence that linear light cone occurs in the $d$-dimensional long-range interacting systems for $\mu>2d+1$ (which determines exponent greater than $3$ in one dimension). 
However, other studies \cite{Tran2020HierarchyInteractions,Eldredge2017FastInteractions} show that beyond that regime either the subpolynomial, polynomial or superpolynomial light cone can occur e.g. while the boundary is calculated using different definitions of operator norms.

Apart from the nature of the coupling, an important factor determining the properties of
lattice systems is the on-site disorder. After the seminal paper of Anderson
\cite{AndersonP.1958} the focus shifted to tight-binding-like models with nearest-neighbor
couplings, where important formal results concerning localization and transport were
obtained \cite{Mott1961, Abrahams1979}. More recently, much interest was devoted to models
of uncorrelated diagonal disorder with power-law long range hopping
\cite{Rodriguez2003AndersonHopping,Dominguez-Adame2004ALattices,DeMoura2005LocalizationInteractions} where transport properties were
characterized via an analysis of localization of states depending on the coupling exponent
and disorder strength \cite{Rodriguez2003AndersonHopping, DeMoura2005LocalizationInteractions}.

In this paper we aim at merging these two aspects and extend the discussion of correlation dynamics to long-range-coupled disordered systems.
We present a complete theory of correlation dynamics in a one dimensional chain of atoms with power-law couplings and strongly disordered on-site energies, based on approximate analytical solutions validated by numerical simulations.
Long-range couplings allow the initially quenched site to communicate with distant atoms immediately, hence we observe immediate spreading of correlations.
We establish a universal triple-phase dynamics of the correlation growth at a given site.
First, the correlations increase like a square of time, then at a certain instant of time there is a change to a growth directly proportional to the time.
Finally, a fixed value (saturation) is achieved.
We are able to provide a fully analytical description of the correlation dynamics.
Consequently, we can find the time needed to achieve some given value of correlation at a given distance $r$ and thus establish analytical formulas for the propagation of correlations.
We show that the correlation dynamics changes from $r(t) \propto t^{1/\mu}$ to $r(t) \propto t^{1/(2\mu)}$ at a certain cross-over time.
A special case represents $\mu=1$, where a strictly linear light cone occurs in the first (ballistic) phase of the motion. However, the continuation of the linear trend is not possible and, at a certain moment of time, the propagation changes to standard diffusion.

The article is organized as follows.
In Section \ref{sec:model} we introduce the investigated system and the theoretical model.
Next, in Section \ref{sec:results} we present and comment on the results of the numerical simulations of the introduced model.
In Section \ref{sec:analytic} we present the results of the analytical approach in the central atom approximation.
Finally, in Section \ref{sec:discussion} we summarize and conclude the work.
\begin{figure}[bt]
    \centering
    \includegraphics[width=0.9\linewidth]{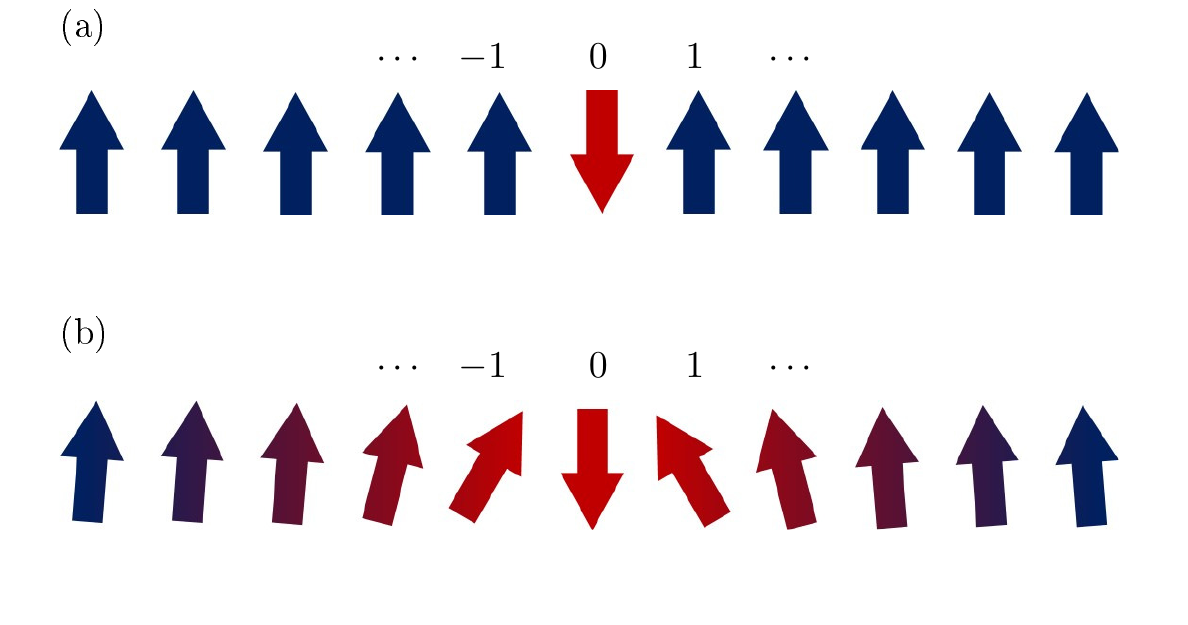}
    \caption{Schematically illustrated physical system --- a chain of uniformly distributed ferromagnetic spins (lattice constant is equal to one) with periodic boundary conditions. 
    (a) When one of the spins is rotated (single local quench), interactions occur in the system. 
    (b) The rotated central spin pulls on the other spins causing excitation to move through the system. 
    In a system with long-range interactions, information about the quench is expected to reach distant spins immediately.}
    \label{fig:system}
\end{figure}
\section{\label{sec:model}Model}
\begin{figure*}
    \centering
    \includegraphics[width=\textwidth]{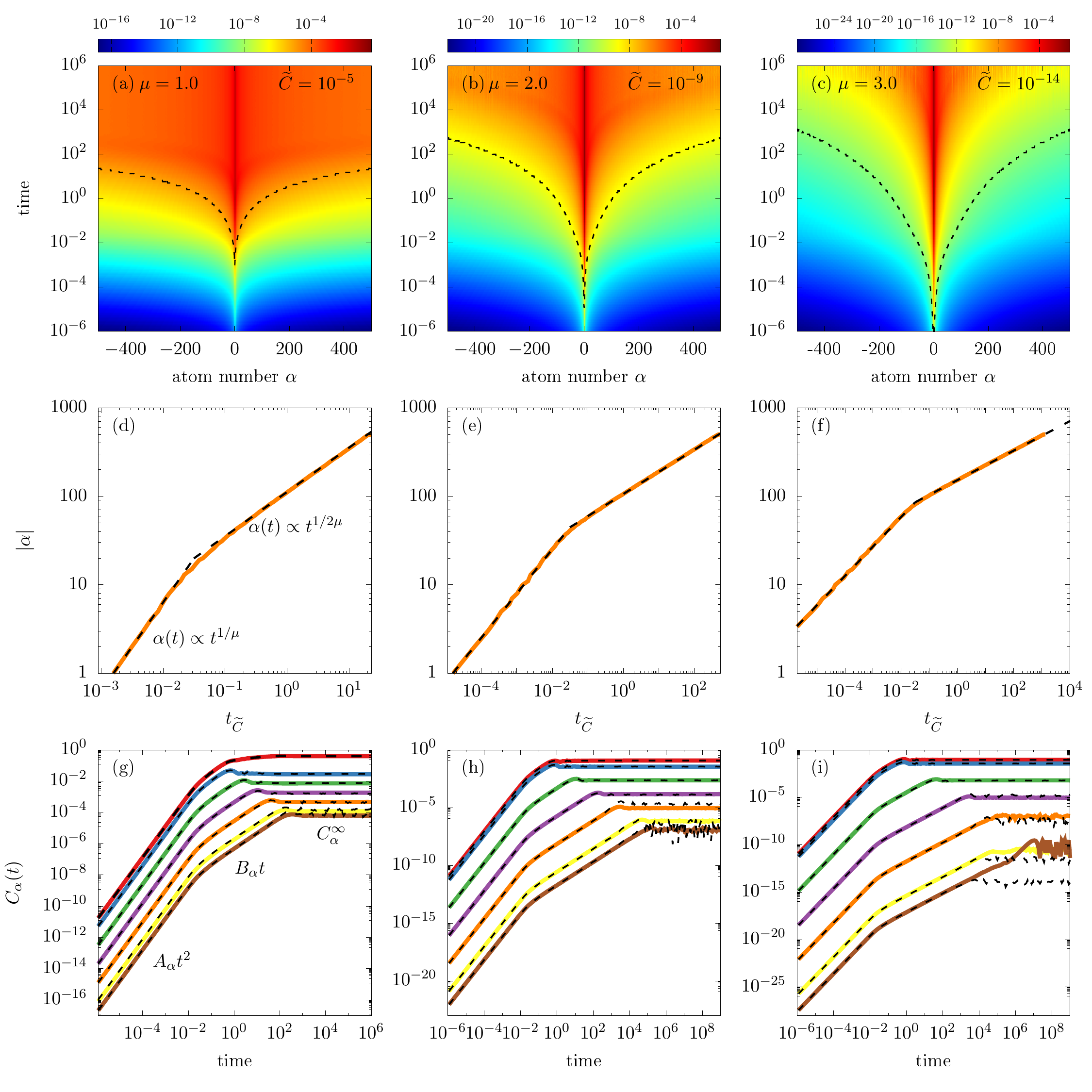}
    \caption{(a-c) Map of time evolution of the absolute value of the correlation function $|C_\alpha(t)|$ (Eq. \eqref{eq:correlator}) for every single node in the system of $N=1001$ spins with disorder strength $W=200$.
    Color maps indicate the values of the correlation function as a function of the site number and time for three different values of the interaction exponent ($\mu=1.0,\ 2.0,\ 3.0$). 
    The dashed lines indicate the propagation fronts defined by $C_\alpha(t)=\widetilde{C}$ (with $\widetilde{C}$ given in the upper right corner of each subfigure).
    The same correlation fronts are shown again in log-log scale in subfigures (d-f), where dashed lines represent analytical results obtained by employing central atom model (see Sec.~\ref{subses:central_atom_model}) and the axes have been swapped to better represent the propagation of the correlation.  \, 
    (g-i) Correlation function $|C_\alpha(t)|$ as a function of time for atoms of $\alpha=0,\ 1,\ 4,\ 16,\ 64,\ 256,\ 500$. Dashed lines here indicate the numerical solution of the central atom model where simulations were performed for $500,000$ disorder realizations.}
    \label{fig:map}
\end{figure*}

In this section, we introduce the system under study and the physical model.
We also present the numerical and analytical methods used to obtain the results.

The system is a chain of $N$ spins distributed on a regular lattice with unit lattice constant. 
The system has periodic boundary conditions.
In this system we consider a local quench leading to a single quasi-particle moving along the chain in both directions (see Fig.~\ref{fig:system}).
The Hamiltonian has the form
\begin{equation}
    H = J\left(\frac{1}{2}\sum\limits_\alpha \epsilon_\alpha\left(1-\Hat{\sigma}_z^\alpha\right) + \sum\limits_{\alpha,\beta} V_{\alpha\beta}\Hat{\sigma}_{+}^{\alpha}\Hat{\sigma}_{-}^{\beta}\right),
     \label{eq:hamiltonian}
\end{equation}
where $\sigma_z^{\alpha}$, $\sigma^\alpha_+ = \sigma_x^\alpha+i\sigma_y^\alpha$ and $\sigma^\alpha_- = \sigma_x^\alpha - i\sigma_y^\alpha$ are single particle spin operators at site $\alpha$.
$J$ sets the overall energy scale,
$J\epsilon_\alpha$ is the on-site energy and $JV_{\alpha\beta}$ is the coupling between spins $\alpha$ and $\beta$.
The dimensionless energies $\epsilon_\alpha$ (in units of $J$) are uncorrelated random variables uniformly distributed on the interval $[-W/2,W/2]$, where the parameter $W$ determines the strength of disorder.
The inter-spin coupling $V_{\alpha\beta}$ has a power-law character,
\begin{equation}
    V_{\alpha\beta} =\left\{ \begin{array}{lrr}
        \dfrac{1}{|\alpha-\beta|^\mu} & \mathrm{for} & \alpha\neq\beta, \\
         0 & \mathrm{for} & \alpha=\beta, 
    \end{array} \right.
\end{equation}
where $|\alpha-\beta|$ is the distance between the spins $\alpha$ and $\beta$.

The central spin ($\alpha=0$) is initially flipped (local quench).
This leads to spreading of the single spin excitation through the system carrying the information about the quench to distant atoms.
Quantitatively, an amount of information shared by the spin $\alpha$ with the initially quenched site is given by the two-point correlation function, which can be measured in experiments,
\begin{equation}
    C_\alpha(t) = \Big\langle\langle \Hat{\sigma}_z^\alpha(t) \Hat{\sigma}_z^0(t) \rangle - \langle \Hat{\sigma}_z^\alpha(t) \rangle \langle \Hat{\sigma}_z^0(t) \rangle \Big\rangle_\mathrm{dis.},
    \label{eq:correlator}
\end{equation}
where $\langle ... \rangle_\mathrm{dis}$ stands for the average over disorder realizations and $\langle ... \rangle$ denotes the quantum mechanical average.
The  operator $\Hat{\sigma}_z^\alpha$ acts on the localized basis state according to,
\begin{equation}
    \Hat{\sigma}_z^\alpha \ket{\beta} = \left\{\begin{array}{rlc}
        -\ket{\beta} & \mathrm{for} &\alpha=\beta, \\
        +\ket\beta & \mathrm{for}&\alpha\neq\beta,
    \end{array}\right.
    \label{eq:spin_operator}
\end{equation}
where in the Schr\"odinger picture, we describe the state vector as
\begin{equation}
    \ket{\Psi} = \sum_\alpha a_\alpha(t) \ket{\alpha},\quad \ket{\alpha} =  \sigma_+^\alpha\ket{\textrm{fm}},
    \label{eq:psi}
\end{equation}
where $\ket{\textrm{fm}}$ represents a system before quench having the ferromagnetic order of spins, $a_\alpha(t)$ are time dependent coefficients of expansion of the system into localized states basis.

Using  Eqs.~\eqref{eq:correlator}, \eqref{eq:spin_operator} and \eqref{eq:psi} it is straightforward to see that
\begin{equation}
    C_\alpha(t) = \left\{\begin{array}{lcc}
        \phantom{+}4|a_0(t)|^2|a_\alpha(t)|^2 & \mathrm{for} & \alpha\neq 0,\\
        -4|a_0(t)|^2(1-|a_0(t)|^2) & \mathrm{for} & \alpha=0. \\
    \end{array}\right.
    \label{eq:formula_for_correlator_by_occupation}
\end{equation}
The correlation function can be then expressed by the occupations of the involved sites. We find the time evolution of the occupations by exact numerical diagonalization of the Hamiltonian \eqref{eq:hamiltonian}, which allows us to compute the correlation function $C_\alpha(t)$ using Eq.~\eqref{eq:formula_for_correlator_by_occupation}. Simulations were done for 25 million disorder realizations. We took the benefit of periodic boundary conditions by obtaining $N$ realizations of disorder from a single diagonalization by arbitrarily choosing the initially rotated spin.

\section{\label{sec:results}Numerical results}
In this section we present the results obtained by a numerical solution of the model described in Sec.~\ref{sec:model}.

\begin{figure*}[tb]
    \centering
    \includegraphics[width=\linewidth]{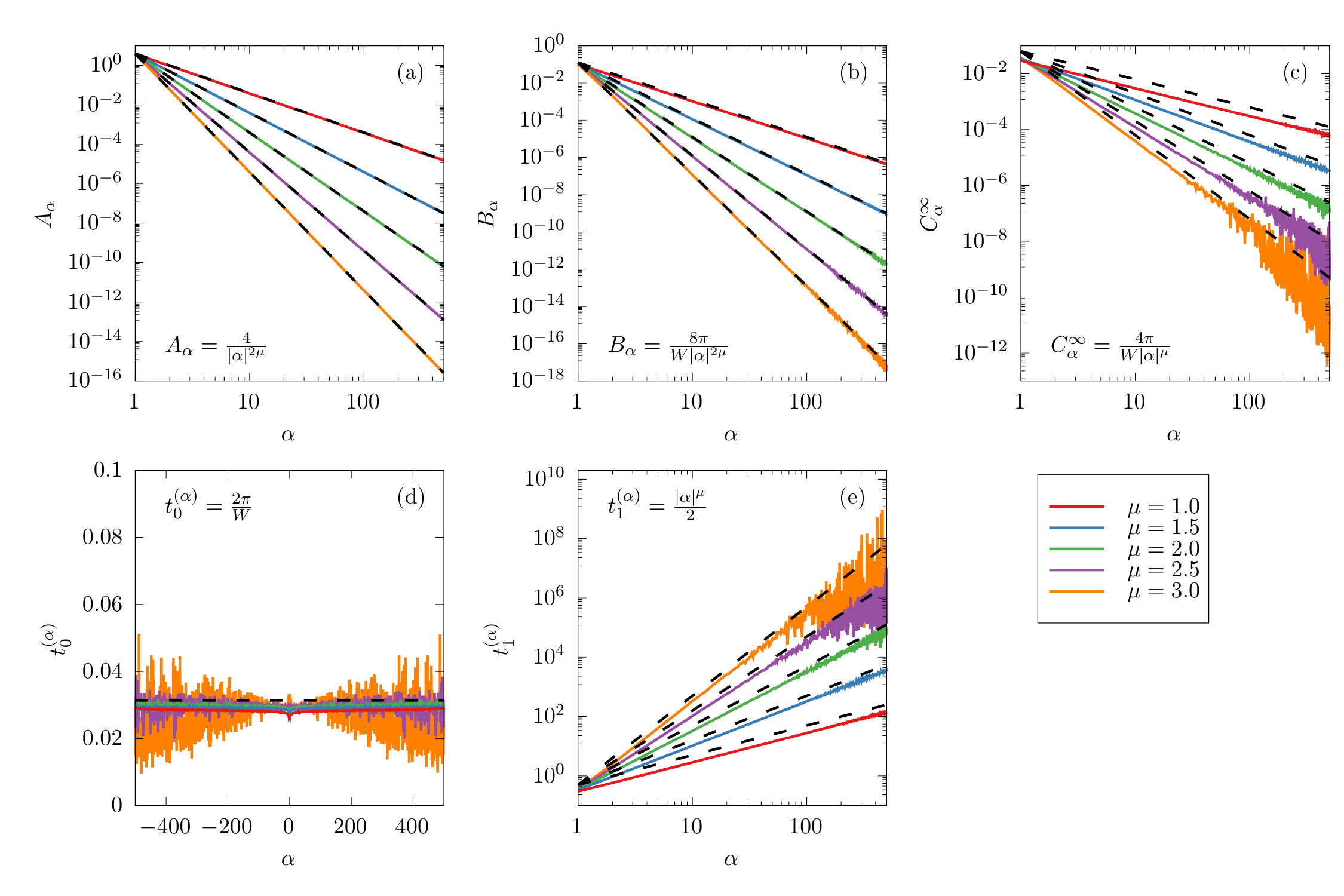}
    \caption{Parameters of the evolution of correlations as a function of the distance from the central atom for several values of exponent the $\mu=1.0,\ 1.5,\ 2.0,\ 2.5,\ 3.0$. In each subfigure, black dashed lines indicate the results obtained from the approximate analytical solution (see Sec.~\ref{sec:analytic}), described by formulas given in the panels. (a) Proportionality factor $A_\alpha$ for the quadratic regime, see Eq.~\eqref{eq:ballistic}; (b) proportionality factor $B_\alpha$ for the linear regime, see Eq.~\eqref{eq:diffusive}; (c) saturation level $C_\alpha^\infty$, see Eq.~\eqref{eq:saturation}; (d) crossover time $t_0^{(\alpha)}$ between quadratic and linear regime; (e) crossover time $t_1^{(\alpha)}$ between linear and saturating regime.}
    \label{fig:evolution_parameters}
\end{figure*}
\subsection{Time Evolution of Correlations \label{subsec:time_evolution_of_correlations}}
Fig.~\ref{fig:map} shows the correlation map for the system of $N=1001$ spins and disorder strength $W=200$ as a function of time, for selected values of the exponent $\mu=1.0,\ 2.0,\ 3.0$.
One can see that the correlations increase with time and saturate at a certain level after a certain time that increases with the node number.
We do not observe any decay of correlations at longer times.
From the correlation dynamics for individual spins we find the propagation front for a pre-defined correlation value $\widetilde{C}$, i.e. a curve in the $\alpha- t$ plane representing the times $t$ at which the correlation at the node $\alpha$ reaches the value $\widetilde{C}$.
These fronts are indicated in Fig.~\ref{fig:map}(a-c) by the dashed lines for the indicated correlation value. 
They are also depicted in Fig.~\ref{fig:map}(d-f) where one can see their power-law nature, changing however towards a decreasing exponent from a certain node number.

To better understand the correlation dynamics, in Fig.~\ref{fig:map}(g-i) we show the time dependence of the correlation for a few nodes ($\alpha = 0,\ 1,\ 4,\ 16,\ 64,\ 256,\ 500$).
For strong disorder, the increase in correlation occurs in qualitatively the same way for each spin, namely one observes three phases of dynamics.

At first, correlations increase as a quadratic function of time,
\begin{equation}
    C_\alpha(t) = A_\alpha t^2, \quad \textrm{for $t<t_0^{(\alpha)}$},
    \label{eq:ballistic}
\end{equation}
where $A_\alpha$ is the appropriate proportionality factor. 
This $\propto t^2$ evolution is a fundamental property of quantum systems derived from the short-time perturbation theory (see e.g.~Ref.~\cite{Colmenarez2020LiebChain}).
Then, at a certain crossover time $t_0^{(\alpha)}$, the time dependence changes to linear,
\begin{equation}
    C_\alpha(t) = B_\alpha t, \quad \textrm{for $t_0^{(\alpha)}<t<t_1^{(\alpha)}$},
    \label{eq:diffusive}
\end{equation}
with $B_\alpha$ as proportionality factor.
Finally, at $t=t_1^{(\alpha)}$ the correlator reaches a constant value (saturates),
\begin{equation}
C_\alpha(t) = C_\alpha^{\infty}, \quad \textrm{for $t_1^{(\alpha)}<t$}.
\label{eq:saturation}
\end{equation}
From the condition of continuity of $C_\alpha(t)$ the crossover times can be expressed by
\begin{equation}
    t_0^{(\alpha)} = B_\alpha/A_\alpha
    \label{eq:t0_continuity relation}
\end{equation}
and
\begin{equation}
    t_1^{(\alpha)} = C_\alpha^\infty/B_\alpha.
    \label{eq:t1_continuity relation}
\end{equation}
The same dynamics was observed for mean squared displacement of a single excitation in the same model (for $\mu=1$) presented in Ref.~\cite{Kawa2020DiffusionCoupling}.

We found the dependence of the dynamical parameters on the site index by fitting  power-law functions to the subsequent regimes of correlation dynamics described by Eq.~(\ref{eq:ballistic}--\ref{eq:saturation}).
The dynamical parameters $A_\alpha$, $B_\alpha$ and $C_\alpha^{\infty}$ appear to be power law functions of the distance $|\alpha|$ with integer or rational exponents.
Both $A_\alpha$ and $B_\alpha$ are inversely proportional to $|\alpha|^{2\mu}$ as it is depicted in Fig.~\ref{fig:evolution_parameters}(a,b). 
The first cross-over time $t_0^{(\alpha)} \equiv t_0$ is roughly distance-independent except for small values of $|\alpha|$ [Fig.~\ref{fig:evolution_parameters}(d)].
As a function of the site number, the saturation level decreases  as $1/|\alpha|^\mu$ (see Fig.~\ref{fig:evolution_parameters}(c)).
The noise in Fig.~\ref{fig:evolution_parameters}(c-e), apparent especially for the high magnitude of exponent $\mu$ and distant sites, comes from the unavoidably insufficient number of disorder realizations. Using a simple resonance-counting argument, one can notice that the number of sites resonant to the central site (in the first order approximation) is proportional to $V/W$, hence for the considered system of $N=1001$ spins, $W=200$ and $\mu=3.5$ one would need a number of disorder realizations at least on the order of $10^{11}$ to allow reach one resonant case on average for distant nodes.

\subsection{Central Atom Approximation \label{subses:central_atom_model}}
In this subsection we introduce the central atom approximation in which analytic solution, well approximating the full model, becomes available \cite{Kawa2020DiffusionCoupling} (see. Sec.~\ref{sec:analytic}).

When the disorder is strong, the coupling can be treated as a perturbation in the Hamiltonian.
In the first-order approximation to the evolution one includes only the couplings between all the spin and the central one.
Thus, the information about the quench is carried directly to the distant spins without the involvement of intermediate jumps.
The correlation dynamics in the central atom approximation reproduces the dynamics of the full model in the strong disorder regime.
The dashed black lines in Fig.~\ref{fig:map}(g-i) represent the simulation results of the central atom approximation, which perfectly match the data of the full model for short-range and moderate sites, although it reveals a discrepancy for the distant sites. This can be however considered as a numerical error coming from too few disorder realizations, as explained above.

The central atom approximation means that each spin can be viewed as the nearest neighbor of the central spin. 
Indeed, in the model with nearest-neighbor couplings, the dynamics of the nearest spins ($|\alpha|=1$) matches the one we found, i.e. triple phase $t^2 \rightarrow t \rightarrow \mathrm{const}$. 
The dynamics of higher order spins changes by increasing the exponent in the time dependence. 
And so, for $|\alpha|=2$ the dynamics of the nearest-neighbor model come through $t^4 \rightarrow t^2 \rightarrow \mathrm{const}$.

\section{\label{sec:analytic}Approximate analytical solution}
The central atom model allows us to find analytical expressions for the correlation function and dynamical parameters by using the theory presented in Ref.~\cite{Kawa2020DiffusionCoupling}. Here we present a more straightforward approach that leads to the same analytical formulas.
We find analytical expressions for the correlation function, the dynamic parameters as a function of the distance from central site, the disorder strength and the exponent $\mu$. This allows us to find the time of reaching a given value of correlation for a given spin and determine the existence of the light cone in the model.

In the high disorder regime the survival occupation of the central site $|a_0(t)|^2\approx 1$.
We can then write the approximate formula for the correlator \eqref{eq:formula_for_correlator_by_occupation}
\begin{equation}
    C_\alpha(t) = 4\langle |a_\alpha(t)|^2\rangle_\mathrm{dis.} \quad \mathrm{for}\quad W\gg 1, \quad \alpha\neq0.
    \label{eq:correlator_simplified_only_one_occupation}
\end{equation}

\subsection{Solution of Two-Spin Model}
\begin{figure}[tb]
    \centering
    \includegraphics[width=\linewidth]{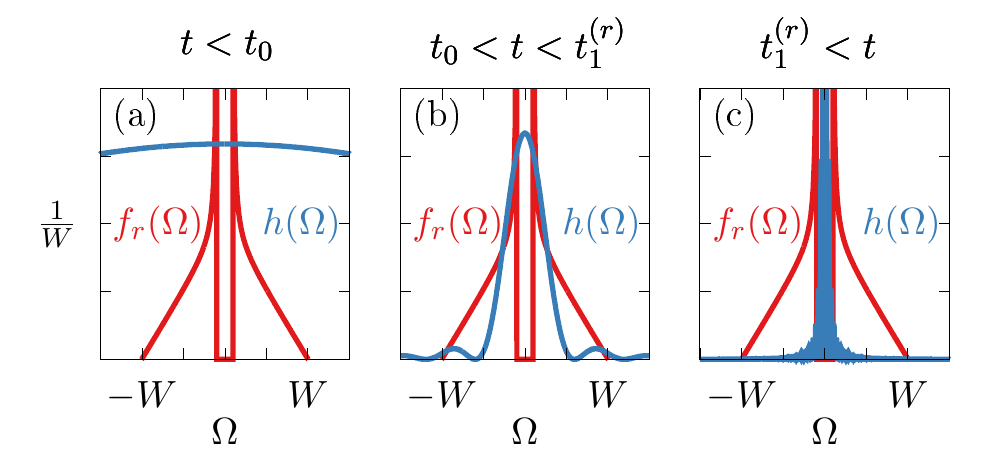}
    \caption{Schematic plot of the probability distribution of eigenenergies separation: (a) for the first phase of motion ($C_\alpha(t)\propto t^2$); (b) for the second phase of motion ($C_\alpha(t)\propto t$); (c) for the saturation phase ($C_\alpha(t) = C_\alpha^\infty$)}
    \label{fig:distribution}
\end{figure}
Since only direct jumps from the central site to distant ones are important (central atom model) in the leading order one can neglect also the presence of other spins to calculate accurately the occupation of the distant site. 
Let us now find these occupations of the individual sites assuming a two-site model.
One considers a system consisting of only two spins, i.e., the central one and the one with index $\alpha$.
The Hamiltonian corresponding to system having two spins of uniformly distributed on-site energies $\epsilon_0$ and $\epsilon_\alpha = \epsilon_0 + \epsilon$ has the matrix form of
\begin{equation}
    H = \left(\epsilon_0 + \frac{1}{2}\epsilon\right)\mathbb{I} + \frac{1}{2}\begin{pmatrix}-\epsilon & 2V_\alpha \\  2V_\alpha & \epsilon \end{pmatrix},
\end{equation}
where $\mathbb{I}$ is $2\times 2$ unit matrix and $V_\alpha = 1/|\alpha|^\mu$ is the coupling to $\alpha$-th spin.
One can easily diagonalize the above Hamiltonian.
The eigenenergies $E_{\pm}$ and corresponding eigenvectors $\ket{\pm}$ are
\begin{gather}
    E_\pm = \epsilon_0 + \frac{1}{2}\epsilon \pm \dfrac{1}{2}\Omega,
    \label{eq:eigenenergies}
\end{gather}
\begin{gather}
    \ket{+} = \begin{pmatrix} \sin\left(\theta/2\right) \\ \cos\left(\theta/2\right) \end{pmatrix},\quad \ket{-} = \begin{pmatrix} \cos\left(\theta/2\right) \\ -\sin\left(\theta/2\right) \end{pmatrix},
    \label{eq:u2}
\end{gather}
where we denote $\Omega = \sqrt{\epsilon^2 + 4V_\alpha^2}$, $\epsilon = \Omega\cos\theta$ and $2V_\alpha = \Omega\sin\theta$.
Then, the time evolution of the amplitude of probability for spin $\alpha$ is given by
\begin{gather}
    a_\alpha(t) = \braket{\alpha}{\Psi(t)} = \mel{\alpha}{e^{-iHt}}{0} = \nonumber \\\sum_{n=\pm} \braket{\alpha}{n}\braket{n}{0}e^{-iE_n t}.
    \label{eq:amplitude_analytic}
\end{gather}
Combining equations (\ref{eq:correlator_simplified_only_one_occupation},\ref{eq:eigenenergies}-\ref{eq:amplitude_analytic}) we get
\begin{gather}
    C_\alpha(t) = 4V_\alpha^2 \left\langle\dfrac{\sin^2\left(\Omega t/2\right)}{\left(\Omega/2\right)^2}\right\rangle.
    \label{correlation_function_average_sine_squared}
\end{gather}
The average over disorder realizations can be obtained by the integration with probability density function of $\epsilon$. 
Since $\epsilon$  is a difference between two uniformly distributed random variables on an interval $[-W/2,W/2]$, its probability density function is the triangle function
\begin{gather}
f_0(\epsilon) = \left\{ \begin{array}{lc}
\left(W-|\epsilon|\right)/W^2,  & \epsilon \in [-W,W];\\ 0,  & \mathrm{otherwise.}
\end{array} \right.
\label{eq:correlator_average_with_distribution}
\end{gather}
Then, the correlation is
\begin{gather}
    C_\alpha(t)= 4V_\alpha^2 \int_{-W}^W d\epsilon f_0(\epsilon) \dfrac{\sin^2\left(\Omega(\epsilon)t/2\right)}{\left(\Omega(\epsilon)/2\right)^2}.
    \label{eq:correlation_function_integral_1}
\end{gather}
By changing the integration variable to $\Omega$, we obtain
\begin{gather}
    C_\alpha(t) = 4V_\alpha^2 \int d\Omega f_\alpha(\Omega) \dfrac{\sin^2\left(\Omega t/2\right)}{\left(\Omega/2\right)^2},
    \label{eq:correlation_function_integral_2}
\end{gather}
where
\begin{gather}
    f_\alpha(\Omega) = f_0(\sqrt{\Omega^2 - 4V_\alpha^2}) \dfrac{|\Omega|}{\sqrt{\Omega^2-4V_\alpha^2}}
\end{gather}
is the probability density function of $\Omega$ for a given atom $\alpha$, i.e., it depends on the distance from the origin of the system.
For increasing $W$ and $|\alpha|$, $f_\alpha(\Omega)$ tends to $f_0(\epsilon)$.

\subsection{Triple Phase Dynamics}
This allows us to explain the triple phase dynamics of the correlation growth.
First, for the very short time scales the function $h(\Omega) = \sin^2\left(\Omega t/2\right)/(\Omega/2)^2$ in Eq.~(\ref{correlation_function_average_sine_squared}, \ref{eq:correlation_function_integral_1}, \ref{eq:correlation_function_integral_2}) can be approximated as $h(\Omega) \approx t^2$ and the integral of the probability density function is equal to unity (see Fig.~\ref{fig:distribution}(a)).
The correlator follows the form
\begin{equation}
    C_\alpha(t) = \dfrac{4t^2}{|\alpha|^{2\mu}}.
\end{equation}
Then according to Eq.~\eqref{eq:ballistic}, the first dynamical parameter is
\begin{equation}
A_\alpha = 4/|\alpha|^{2\mu}, 
\end{equation}
which is sketched by dashed lines in Fig.~\ref{fig:evolution_parameters}(a) and matches perfectly the numerical data.

Next, for moderate times $t_0<t<t_1$ [Fig.~\ref{fig:distribution}(b)] the function $h(u)$ probes the central part of the distribution but is still relatively broad and therefore insensitive to the narrow central gap (especially for remote nodes).
For increasing time, $h(\Omega)$ tends to be proportional to unnormalized Dirac delta of area of $2\pi t$.
Then Eq.~\eqref{eq:correlator_average_with_distribution} takes the form,
\begin{equation}
    C_\alpha(t) \approx 4 |V_{\alpha 0}|^2\int_{-\infty}^{\infty}d\Omega f_\infty(\Omega) 2\pi t\delta(\Omega) = \dfrac{8\pi}{W|\alpha|^{2\mu}}t,
    \label{eq:integral_diffusive}
\end{equation}
where, according to Eq.~\eqref{eq:diffusive}, the second parameter is
\begin{equation}
    B_\alpha = \dfrac{8\pi}{W|\alpha|^{2\mu}},
\end{equation}
which fits perfectly the data in Fig.~\ref{fig:evolution_parameters}(b).

From the requirement of continuity, Eq.~\eqref{eq:t0_continuity relation}, the crossover time between the first and second phases is
\begin{equation}
    t_0 = \dfrac{2\pi}{W}
\end{equation}
and is the same for all the sites.

Finally, we obtain the saturation level.
When the function $h(u)$ becomes narrow, its peak coincides with the gap inside the distribution $f_\alpha(\Omega)$ [Fig.~\ref{fig:distribution}(c)].
Then the peak does not contribute and only the oscillating tail has a contribution to the integral.
We approximate $h(\Omega)\approx \frac{1/2}{(\Omega/2)^2}$ taking average of the sine squared function in the numerator.
Then the saturation of the correlation is
\begin{align}
    \begin{split}
       C_\alpha^{\infty} &= 8\int_{2V_r}^{\sqrt{W^2+4V_\alpha^2}} d\Omega f_\alpha(\Omega) \dfrac{1/2}{\left(\Omega/2\right)^2}\\ =&\dfrac{4\pi}{W|\alpha|^\mu}\arctan\left(W|\alpha|^\mu\right) - \dfrac{8}{W|\alpha|^{2\mu}}\ln\left[1+\left(\frac{W|\alpha|^\mu}{2}\right)^2\right]\\
       \approx &  \dfrac{4\pi}{W|\alpha|^\mu}, 
    \end{split}
    \label{eq:saturation_integrated_with_distribution}
\end{align}
where the approximate result was obtained by taking $\arctan(x) \approx \pi/2$ for $x\gg1$  and by neglecting the second order terms in $V_\alpha/W$.
The analytical formula from Eq.~\eqref{eq:saturation_integrated_with_distribution}, with dashed lines marked in Fig.~\ref{fig:evolution_parameters}(c), overestimates the numerical results.
However, the deviation from the numerical data decreases as the exponent $\mu$ increases.
This discrepancy may be caused by the inaccuracy of the approximation done in Eq.~\eqref{eq:correlator_simplified_only_one_occupation}, where we assumed the occupation of the central spin, $|a_0(t)|^2$, close to unity. The asymptotic survival probability, $|a_0(\infty)|^2$, can actually subtly but not marginally deviate from unity, the less the higher $\mu$ is. Then, the approximation in Eq.~\eqref{correlation_function_average_sine_squared} is more accurate if the exponent $\mu$ is large.

From the second continuity relation in Eq.~\eqref{eq:t1_continuity relation} we get the second crossover time,
\begin{align}
    t_1^{(\alpha)} = \dfrac{|\alpha|^\mu}{2}.
\end{align}

The above derivations explain the origin of the triple dynamics and find correct values for the dynamical parameters. 

\subsection{Speed of Propagation --- Derivation of the Light Cone \label{subsec:light_cone}}

Now we want to extract the speed of propagation of the correlation from the above derivations. 
For this purpose, we will find the time at which the given spin reaches a certain value of the correlation $\widetilde{C}$. 
The result depends on which phase of the dynamics $\widetilde{C}$ is in for a particular spin.

Thus, if $\widetilde{C}$ is in the first (quadratic) phase [Eq.~\eqref{eq:ballistic}] the time required to reach such a correlation is
\begin{gather}
    t_{\widetilde{C}} = \frac{\sqrt{\widetilde{C}}}{2}|\alpha|^\mu
    \label{eq:t_CA}
\end{gather}
which has to be less than $t_0$. 
On the other hand, if $\widetilde{C}$ is in the second (linear) phase [Eq.~\eqref{eq:diffusive}], the time to reach this correlation value is
\begin{gather}
    t_{\widetilde{C}} = \frac{W\widetilde{C}}{8\pi} |\alpha|^{2\mu}.
    \label{eq:t_CB}
\end{gather}
These dependencies are sketched by dashed lines in Fig.~\ref{fig:map}(d-f) and match perfectly numerical data.
On a doubly logarithmic scale, these are straight lines indicating the power-law dependencies. 
 
The propagation character changes with time. 
In the first phase $|\alpha (t)| \propto t^{1/\mu}$ while in the second phase the dynamics slows down and $|\alpha(t)|\propto t^{1/2\mu}$. 
The limiting time between the two regimes is $t_0$, for which the correlation dynamics changes from $\propto t^2$ to $\propto t$. 
It is independent of the choice of spin number or even exponent, so regardless of the choice of $\widetilde{C}$, the cross-over will be observed at the same point (dependent only on the strength of disorder).\vfil

\section{\label{sec:discussion}Discussion}
The propagation of correlation calculated in Sec.~\ref{subsec:light_cone} varies depending on whether the given correlation value is within the first or second phase of the dynamics. 
This means that the character of propagation changes depending on the phase of the dynamics.
This can be clearly seen in Fig.~\ref{fig:map}(d-f), where the time to reach given correlation $\widetilde{C}$ is shown as a function of distance from the center of the chain.
One can see the change in the slope of the trend from a certain distance. 
An initially ballistic motion ($\alpha(t) \propto t$) for $\mu=1$ changes to a standard diffusion ($\alpha(t)\propto \sqrt{t}$). Ballistic motion implies the existence of a linear light cone i.e., a constant propagation velocity. For $\mu>1$, one can think of a sublinear ``cone''.

One can also ask whether the finite size of the system plays any role. It is particularly concerning for the case of $\mu = 1$, where the saturation level of the correlation (and thus the site occupation) decreases as $1/r$. Therefore the sum of the occupations is divergent and thus one may expect the model to break down at some point. However, in the thermodynamic limit, the saturation phase for very far atoms begins at infinite times since $t_1\propto r \to \infty$. Infinite $t_1$ implies the occupation cannot saturate and remains in second phase of evolution [Eq.~\eqref{eq:diffusive}], which diminishes as $1/r^2$ and the corresponding sum of occupations is convergent.

\section{Conclusions}
We investigated the propagation of correlations in a spin chain after a single local quench in the presence of large disorder and long-range couplings.
The main feature observed in the system is the triple-phase evolution of correlation at each site which results in a change in the propagation trend of the correlation front. 
The ``light cone'' in the strongly disordered system can be strictly linear only in the particular case of $\mu=1$ and only as long as the correlation is in the first phase of motion. For $\mu>1$ the propagation is sub-ballistic in the first phase and becomes sub-diffusive in the second phase of motion.
All the effects observed in the numerical simulations are explained by an analytical model.

\section{Acknowledgments}
Calculations  have  been  partially  carried  out  using resources provided by Wroclaw Centre for Networking and Supercomputing \footnote{\url{https://www.wcss.pl/}}, Grant No. 203.
\bibliography{my,apssamp}
\end{document}